# Controllable Strain-driven Topological Phase Transition and Dominant Surface State Transport in High-Quality HfTe$_5$ Samples

Jinyu Liu[1], Yinong Zhou[1], Sebastian Yepez Rodriguez[1], Matthew A. Delmont[2], Robert A. Welser[1], Nicholas Sirica[3], Kaleb McClure[4], Paolo Vilmercati[4], Joseph W. Ziller[5], Norman Mannella[4], Javier D. Sanchez-Yamagishi[1], Michael T. Pettes[3], Ruqian Wu[1], Luis A. Jauregui[1,*]

[1]Department of Physics and Astronomy, University of California, Irvine, CA 92697, USA

[2]Department of Mechanical and Aerospace Engineering, University of California, Irvine, CA 92697, USA

[3]Center for Integrated Nanotechnologies (CINT), Materials Physics and Applications Division, Los Alamos National Laboratory, Los Alamos, NM 87544, USA

[4]Department of Physics and Astronomy, The University of Tennessee, Knoxville, TN, 37996, USA

[5]Department of Chemistry, University of California, Irvine, CA 92697, USA

* Corresponding email: lajaure1@uci.edu

**Abstract:**

Controlling materials to create and tune topological phases of matter could potentially be used to explore new phases of topological quantum matter and to create novel devices where the carriers are topologically protected. It has been demonstrated that a trivial insulator can be converted into a topological state by modulating the spin-orbit interaction or the crystal lattice. However, there are limited methods to controllably and efficiently tune the crystal lattice and at the same time perform electronic measurements at cryogenic temperatures. Here, we use large controllable strain to demonstrate the topological phase transition from a weak topological insulator phase to a strong topological insulator phase in high-quality HfTe$_5$ samples. After applying high strain to HfTe$_5$ and converting it into a strong topological insulator, we found that the sample's resistivity increased by more than two orders of magnitude (24,000%) and that the electronic transport is dominated by the topological surface states at cryogenic temperatures. Our findings show that HfTe$_5$ is an ideal material for engineering topological properties, and it could be generalized to study topological phase transitions in van der Waals materials and heterostructures. These results can pave the way to create novel devices with applications ranging from spintronics to fault-tolerant topologically protected quantum computers.

**Introduction:**

The concept of topology in condensed matter physics has revolutionized our understanding of the electronic band structure of solid-state materials[1–9]. A topological phase transition (TPT), a continuous quantum phase transition between states with different topological order, can occur by tuning certain physical parameters[2,5,10–14]. As observed in several pioneering experiments, such transitions from trivial insulators to topological insulators (TI) have been realized by tuning the lattice parameters and/or spin-orbit coupling through element substitution[15–17]. While substitutional doping can introduce disorder into the material, the application of strain is a cleaner method for anisotropically tuning the lattice constants in a bulk or microscopic device sample[18,19]. First, in order to better understand TPTs in solid-state systems, and to unleash strain's potential to create novel phases of matter[20], pseudo-magnetic fields[21,22], or manipulate the Berry curvature locally[23,24] to create quantum devices for quantum electronics and spintronics, it is desirable to find a quantum material whose topological properties are sensitive to strain. Here, we demonstrate that $HfTe_5$ is an ideal material for studying TPTs through electrical transport under large and controllable strain.

Since the prediction of their nontrivial band topology[25], transition metal pentatellurides such as $ZrTe_5$ and $HfTe_5$, have shown a range of interesting physical properties including the chiral magnetic effect[26,27], the anomalous Hall effect[28], and the three-dimensional (3D) quantum Hall effect[29,30]. The pentatellurides crystallize in a layered orthorhombic structure with a space group of *cmcm*, as shown in Fig. 1a for the crystal structure of $HfTe_5$. As the building block of a two-dimensional (2D) layer, each $HfTe_3$ trigonal prism is formed by a Te-d dimer and an apical Te-a atom. The trigonal prisms with a one-dimensional (1D) characteristic along the *a* axis are linked by parallel Te-z atoms with zig-zag chains along the *c* axis, assembling a 2D $HfTe_5$ layer in the *ac* plane. The $HfTe_5$ layers are stacked together along the *b* axis through van der Waals interactions to form a 3D layered crystal. The topological nature of as-grown samples has been debated, with different experiments finding that the pentatellurides can be either a weak topological insulator (WTI)[31–34], a strong topological insulator (STI)[27,35,36], or a Dirac semimetal (DSM)[26,37–39]. This inconsistency may be because the pentatellurides lie close to the phase boundary between a WTI and an STI phase[25]. First-principles calculations have shown that the topological character of the pentatellurides can change significantly with slight variations in the lattice parameters[25]. For 3D TIs, the STI phase exhibits topological surface states (TSS) on all of the surfaces, while a WTI phase hosts TSS only on the side surfaces. Electrons in an STI are topologically protected against backscattering, while electrons in a WTI phase are not necessarily protected[4]. A WTI and an STI have distinct topological characteristics and a transition between the two phases are not possible without closing and re-opening the bulk gap. For $HfTe_5$, in the WTI phase the TSS are predicted to exist only on the side surfaces, i.e. (100) and (001) planes, but not on the top and bottom surfaces along the (010) planes. The WTI phase resembles a 2D quantum spin Hall edge states, all stacked along the *b* axis. $HfTe_5$ being at the phase boundary between a WTI and an STI phase is the perfect candidate for studying TPTs driven by strain.

Recently, signatures of a TPT from a WTI to an STI phase in ZrTe$_5$ were revealed by angle-resolved photoemission spectroscopy (ARPES)[33] and electrical transport measurements[40,41]. However, the defining magneto-transport evidence of an STI phase, the dominant contribution of the TSS upon the application of strain, is still missing. Also, less attention has been paid to HfTe$_5$ even though it has a larger spin-orbit coupling than ZrTe$_5$, which potentially could facilitate the formation of band inversion. Here, we present a comprehensive study of the strain-driven TPT from a WTI to an STI phase in high-quality HfTe$_5$ samples from both in-depth first-principles calculations and extensive quantum transport experiments. First, by combining the Density Functional Theory (DFT) calculated phase diagram of HfTe$_5$ and the electrical transport measurements in a small strain range, we show our as-grown HfTe$_5$ samples are in a WTI phase. As we apply a larger strain, we observe the band gap is initially closing, showing dramatic changes in the temperature dependence of the resistivity. As strain is increased further the band gap reopens converting the low-temperature behavior from initially metallic to insulating. With the largest amount of strain (~ 4.5%) applied, the resistivity saturates for temperatures, $T < 10$ K, hinting at a dominant TSS transport as observed in other well-studied 3D TIs[42–45]. In addition, our results from magneto-transport measurements with the magnetic field oriented along both the $b$ axis (out-of-plane direction) and the $a$ axis (in-plane direction, with $I // B$) agree well with a gapped bulk and a gapless TSS of an STI phase under large strain. Our observations of the band gap closing, transition from a metallic to insulating behavior, the saturated resistivity at low temperatures, and TSS dominant magneto-transport under large strain are unambiguous experimental electrical transport signatures of a strain-driven TPT in HfTe$_5$.

**Results:**

We have synthesized belt-like HfTe$_5$ single crystals with a typical length of 1 cm by the chemical vapor transport (CVT) method (see Methods). The full lattice information of our samples, obtained from the refinement of single crystal diffraction is included in Section I of the Supplementary Information (SI). The crystallographic $a$ and $c$ axes are along the in-plane length and width directions respectively. The electronic structure is measured by ARPES. Measurements were carried out on an in-situ cleaved (010) surface, where high symmetry cuts along the XΓX ($k_a$) and YΓY ($k_c$) axes are shown in Figs. 1b and 1c, respectively. Here, prominent hole-like dispersion near the Brillouin zone center is seen in the vicinity of the Fermi level ($E_F < E_{BE} < 0.4$ eV), where $E_{BE}$ is the binding energy. A single Fermi surface pocket is detected at the $E_F$. The overlaid green lines correspond to the first-principles calculation results as discussed below and plotted here to show the agreement between ARPES experiments and calculations. While no photon energy dependence was performed in this study, the similarity of the measured electronic structure with literature suggests bands dispersing towards $E_F$ to be bulk-derived[34,46].

We characterize the basic electrical transport properties of HfTe$_5$ by measuring "free-standing" samples as described in the Methods. For all the electrical transport experiments, the current is applied along the $a$ axis. A representative temperature dependence of the resistivity measured by the four-probe method, $\rho_{xx}$, is shown in Fig. 1d. As the sample is cooled down, $\rho_{xx}$ decreases slightly between 180 K < $T$ < 300 K. As the sample is cooled further, $\rho_{xx}$ increases drastically for $T$ < 180 K until a peak in $\rho_{xx}$ is reached at $T_p \sim$ 70 K. For $T < T_p$ the sample behaves as a metal (decreasing $\rho_{xx}$ with decreasing $T$). Our measured $T_p \sim$ 70 K agrees well with previously studied HfTe$_5$ samples grown by CVT[30]. Recent ARPES experiments attributed $T_p$ to a temperature-induced Lifshitz transition, where the chemical potential gradually shifts from the valence band to the conduction band of the gapped Dirac cone as the temperature is reduced[46,47], with a hole-dominated transport for $T > T_p$ and electron-dominated transport for $T < T_p$, while the chemical potential is in the gap for $T_p$, as previously reported[30,48] and measured in our samples (Supplementary Fig. S3). One of the explanations of the Lifshitz transition is the reduced lattice constant with the reduced temperature, where the lattice constant $b$ was shown to reduce by 0.3% when cooling down from 300K to 4K[46]. Other plausible explanations are based on Dirac polarons[49] and thermodynamically induced carriers[50], but the origin is still under debate.

We measure the magnetotransport properties of our samples. The $\rho_{xx}$ and the transverse Hall resistivity, $\rho_{xy}$, at $T$ = 1.5 K are plotted vs. the perpendicular magnetic field ($B$) in Fig. 1e. $\rho_{xx}$ displays clear Shubnikov-de Haas (SdH) oscillations starting from $B$ = 0.2 T. The SdH oscillations have a typical frequency of 1.22 T, corresponding to a small Fermi surface cross-sectional area (for $B$ // $b$ axis) $\sim 1.16 \times 10^{-4}$ $Å^{-2}$, extracted by the Onsager relation. The $\rho_{xy}$ vs. $B$ is linear for $B$ < 0.2 T and displays quantum oscillations similar to those seen in $\rho_{xx}$. From the measured $\rho_{xy}$ vs. $B$, we extract an electron carrier density of $3.76 \times 10^{16}$ cm$^{-3}$ (extracted from a single band model) and electron mobility of $755,000$ cm$^2$ V$^{-1}$ s$^{-1}$, which is the highest mobility reported in HfTe$_5$ samples. Beyond the Landau level, $n$ = 1, the quantum oscillations of $\rho_{xy}$ develop an evident Hall plateau, while $\rho_{xx}$ reaches a minimum, demonstrating we observe the 3D quantum Hall effect in our samples, as observed in previous studies[29,30]. These results are consistent with the electron transport being dominated by massive Dirac fermions.

We further investigate the topological properties of HfTe$_5$ by first-principles calculations. The lattice constant and band gap are dramatically influenced by different exchange-correlation functionals. We test four different functionals: standard Perdew-Burke-Ernzerhof (PBE)[51], PBE-D3 with a van der Waals (vdW) correction[52], optB86b-vdW (a modified vdW-DF functional)[53], and the strongly constrained and appropriately normed (SCAN)[54] with the revised Vydrov-van Voorhis (rVV10)[55] (see Table S4 in the SI). The SCAN meta-generalized gradient approximation can accurately treat short- to intermediate-range vdW interactions. We conclude that SCAN with rvv10 vdW correction can better describe the TPT of the pentatellurides with Te-Te bonds in the range of 2.7 - 4 Å. The SCAN-rVV10 approach predicts HfTe$_5$ with no strain applied is in a WTI phase with the Dirac gap at the $\Gamma$ point and the $Z_2$ indices (0;010) close to the phase transition point (see Sections V & VI of the SI).

Next, we study HfTe$_5$ under uniaxial strain along the *c* axis. Figs. 2 a-c shows the band structure along X-$\Gamma$-Y high symmetry lines, where X' is the point along X-$\Gamma$ direction. With a -1% strain applied along the *c* axis the band gap at the $\Gamma$ point is opened (Fig. 2a). Under tensile strain the band gap closes (Fig. 2b) and reopens (Fig. 2c) with a band inversion between the *p* orbitals of Te-d atoms and Te-z atoms (labeled in Fig. 1a). The evolution of the band gap at the $\Gamma$ point is depicted in Fig. 2d, where the gap closing point needs a minimal tensile strain, as indicated by the red dashed line. Based on these results, HfTe$_5$ can transition from a WTI to an STI phase by applying tensile strain along the *c* axis. Similarly, the system can also transition from a WTI to an STI phase by applying compressive strain along the *a* axis, as shown in Fig. 2e. At the $\Gamma$ point the band gap displays a minimum as a function of strain, across the phase transition (Supplementary Fig. S6 and Fig. S7). Our theoretical results confirm HfTe$_5$ is in a WTI phase with no strain applied and a TPT to an STI phase is predicted with the application of tensile strain along the *c* axis.

In order to experimentally confirm that our pristine HfTe$_5$ samples are in the WTI phase, we apply small strain along either the *a* or *c* axis independently. Two different samples $S_a$ and $S_c$ are glued on the side wall of a piezo stack actuator with the *a* axis parallel or perpendicular to the strain direction, as shown in the inset of Fig. 2f. A small strain of up to +/- 0.13% can be continuously applied and monitored (see Section VIII of the SI). Fig. 2f shows the $\rho_{xx}$ *vs.* strain ($\epsilon$) for both of the samples measured simultaneously at *T* = 70 K. We choose *T* = 70 K because the chemical potential is at the band gap and the change of resistance is related to changes in the gap size. For Sample $S_a$ with strain applied along the *a* axis ($\epsilon_a$), $\rho_{xx}$ increases monotonically with increasing $\epsilon_a$, indicating that $\epsilon_a$ favors a band gap-opening. While, for Sample $S_c$ with strain applied along the *c* axis ($\epsilon_c$), $\rho_{xx}$ decreases monotonically with increasing $\epsilon_c$, indicating that $\epsilon_c$ favors a band gap-closing. Our ARPES measurements, electrical transport measurements, and the first-principles calculations agree that we are probing the bands around the $\Gamma$ point. The fact that the topological band gap closes with increasing tensile $\epsilon_c$ (or compressive $\epsilon_a$) unambiguously demonstrates that our HfTe$_5$ samples are intrinsically in the WTI phase[32–34,41,46]. The small band gap of HfTe$_5$ at the $\Gamma$ point, the chemical potential at the band gap (*T* = 70 K), and the opposite trend of $\rho_{xx}$ vs. $\epsilon_a$ when compared to $\rho_{xx}$ vs. $\epsilon_c$ are in good agreement with our first-principles calculations that strain can modify the band gap size. However, our piezo stack actuator cannot be used to apply strain larger than 0.13% therefore we have not observed a full gap closening (minimum of $\rho_{xx}$ vs. $\epsilon$) within this range at *T* = 70 K. By cooling down the sample below $T < 70$ K we observe $\rho_{xx}$ can be reduced by increasing $\epsilon_c$ with a smaller effectivity (Supplementary Figure 9). At $T = 16$ K positive $\epsilon_c$ (tensile strain) becomes ineffective at reducing $\rho_{xx}$ as $\rho_{xx}$ reaches a minimum. This minimum of $\rho_{xx}$ is related to the full closing of the band gap. Our small strain experiment agrees well with our first-principles calculations that only a small strain is needed to close the gap.

To achieve high ($\epsilon_c > 0.13\%$) and uniformly distributed strain on our samples, a different apparatus, "the bending station", inspired by a similar one used for photoemission measurements elsewhere[56,57], is adapted here for electrical transport measurements, as seen in Fig. 3a. To the best of our knowledge, it is the first time that "the bending station" has been used for electrical transport experiments. The details of our experimental design and finite element analysis (FEA) simulation are described in Methods. In short, an air-annealed titanium (Ti) beam is used as the sample substrate and secured against a sapphire bead by four screws, as seen in Figs. 3b i and ii. We use Ti because of the similar thermal expansion coefficient with HfTe$_5$. We apply strain by controlling the position of the four screws. This setup can prevent the sample from buckling as the sample is glued at the center of the Ti beam top surface. An FEA simulation of the strain distributions for $\epsilon_y$ and $\epsilon_x$ along the length and width directions of the beam with a medium-high bending radius is depicted in Figs. 3b iii and iv respectively. To drive our HfTe$_5$ samples from WTI to STI, compressive strain along the $a$ axis or tensile strain along the $c$ axis is required. A highly uniform strain distribution is ensured within the sample area on the top surface of the beam to apply tensile strain along the $c$ axis. On the other hand, applying large compressive strain along the $a$ axis can only be done by gluing it on the bottom of the beam, which would be not uniform and create buckling in the sample. We approximate the upper bound of strain applied to the sample by $\epsilon = t/(2R)$, where $\epsilon$ is the strain, $t$ is the thickness of the beam ($t = 1$ mm), and $R$ is the bending radius. We find that the extracted $\epsilon$ is in good agreement with $\epsilon_y$ from the FEA simulations.

The temperature dependence of $\rho_{xx}$ shows a remarkable evolution with strain as seen in Fig. 3c (for clarity Fig. 3d is plotted on a log-log scale). For our as-glued samples with no intentional strain applied, $\epsilon_0$, no difference is observed in the measured $\rho_{xx}$ vs. $T$ when compared to the "free-standing" samples. For $T < 70$ K, $\rho_{xx}$ is continuously decreasing, in contrast to the sample pasted on the piezo stack. Based on the radius analysis, we find an insignificant strain of $\epsilon_0 \sim 0.04\%$ which may be induced when securing the Ti beam against the sapphire bead by four screws. As the beam is further bent by tightening the four screws, the sample is strained more along the $c$ axis ($\epsilon_c$). By applying a small strain $\epsilon_1 \sim 0.26\%$, we notice a reduction of $\rho_{xx}$ by 32% at $T = 70$ K (when the chemical potential is at the band gap). This reduction of $\rho_{xx}$ with $\epsilon_c$ agrees with our single piezo stack actuator measurements. Applying $\epsilon_1$ by using the "bending station" we observe a larger change of $\rho_{xx}$ when compared with a 19% reduction (with $\epsilon_c = 0.13\%$) by using the single piezo actuator. For 15 K < $T$ < 150 K, $\rho_{xx}$ is much reduced when compared to the measurement under $\epsilon_0$ and this result points towards a reduction of the sample band gap to the point to be closed for $T \sim 20$ K. Such reduction of $\rho_{xx}$ caused by strain $\epsilon_1$ diminishes as the temperature is increased, and becomes negligible for $T$ > 200 K. While for $T$ < 20 K we observe a different trend than the one observed in free-standing samples, we observe $\rho_{xx}$ increases with decreasing T, as the sample band gap has increased or reopened by $\epsilon_1$. As observed in the sample pasted on the piezo actuator. By increasing strain further, $\epsilon_2 \sim 2.3\%$, we observe a more obvious change of $\rho_{xx}$ vs. $T$. First, $\rho_{xx}$ vs. $T$ shows a more dominant insulating behavior. Indicating $\epsilon_2 \sim 2.3\%$ along the $c$ axis is enough to change the electronic structure significantly. As one can see from the high-temperature range, the metallic to semiconducting crossover temperature is $T \sim 240$ K, much higher than the crossover observed in the unstrained samples $\sim 200$ K. Second, at $T = 1.5$ K the resistivity is dramatically enhanced, and it becomes $\sim 2,000\%$ larger than that under $\epsilon_0$. This indicates that the band gap is reopening with increasing strain as shown by $\epsilon_2$ for $T$ < 20 K. We increase strain further to $\epsilon_3 \sim 4.5\%$ and the measured $\rho_{xx}$ vs. $T$ shows an even more insulating behavior. Now, instead of showing crossovers between semiconducting and metallic behaviors, $\rho_{xx}$ increases monotonically as the temperature decreases. Compared with the $\epsilon_0$ conditions, the $\rho_{xx}$ at $T = 1.5$ K has increased by more than two orders of magnitude (24,200%). Interestingly, $\rho_{xx}$ shows a saturation for $T$ < 10 K, as seen in Fig. 3d. This saturation in the $\rho_{xx}$ for $T$ < 10 K is reminiscent of the topological surface state dominant transport previously observed in 3D STIs[42–45].

The evolution of $\rho_{xx}$ vs. $T$ under different strains can be explained by the strain-induced TPT. From the Arrhenius analysis of $\rho_{xx}$ vs. 1/$T$ shown in Fig. 3e we extract the evolution of the bulk band gap. First, for 80 K < $T$ < 120 K for the unstrained sample, a thermal activation energy gap $\Delta \sim 33.6$ meV is extracted, consistent with the Dirac cone gap previously observed by ARPES in HfTe$_5$[34,46]. By applying strain $\epsilon_1 \sim$ 0.26% the extracted gap is reduced to 27.7 meV, which confirms the closing of the band gap with strain as predicted by our DFT calculations. Under large strains $\epsilon_2 \sim 2.3\%$ and $\epsilon_3 \sim 4.5\%$, the insulating behavior of the resistivity starts at higher $T$'s, closer to room temperature. For 170 K < $T$ < 250 K and under $\epsilon_3$ conditions, we obtain $\Delta \sim 38.6$ meV. The increase of the extracted energy gap and the increase of the starting $T$ of the insulating behavior for the $\epsilon_2$ and $\epsilon_3$ cases indicate a reopening of the band gap, driven by large strain. For $T$ < 10 K, strain $\epsilon_3$ shows a saturation of $\rho_{xx}$ as shown in Fig. 3f. Similar $\rho_{xx}$ vs. $T$ has been observed in 3D STIs in previous studies[42–45].

To further investigate the strain-induced TPT, we have performed magnetotransport measurements under different strains. Fig. 4a shows $\rho_{xx}$ vs. $B$. Our sample shows strong SdH oscillations under all strain conditions. To study the SdH oscillations, we plot the oscillatory part ($\Delta\rho_{xx}$) after background subtraction in Fig. 4b. We observe the frequency of the SdH oscillations under strain $\epsilon_2$ (1.40 ± 0.2 T) is 16% larger than that of the unstrained $\epsilon_0$ case (1.21 ± 0.2 T). The 2D carrier density extracted from SdH oscillations by $n_{2D} = F/\Phi_0$, (where $F$ is the oscillating frequency and $\Phi_0$ is the magnetic flux quantum) increases from 5.85 ×10$^{10}$ cm$^{-2}$ for $\epsilon_0$ to 6.77 ×10$^{10}$ cm$^{-2}$ for $\epsilon_2$. An increase in the carrier density is contradictory with the increased bulk band gap by strain, if we only consider transport in the bulk. One plausible explanation is that, under strain $\epsilon_2$, the TSS start to contribute to the total conduction and the probed SdH oscillations are from the TSS. The emergence of the surface state transport is in line with the topological phase transition from a WTI to an STI phase driven by strain. Fig. 4b, also clearly displays a systematic shift of the SdH oscillations as strain is increased. This phase shift could be related to the change in the oscillation frequency and possibly to a phase difference. We note two interesting facts about the oscillations in the high field range ($B > 1$ T). First, the oscillating amplitude is anomalously enhanced near the 1$^{st}$ Landau level ($n = 1$). Second, the interval between the last peak and valley is smaller in $1/B$ than those for $n > 2$. Both indicate the electronic bands are greatly influenced at a higher field ($B > 1$ T) when approaching the quantum limit, which is expected given the large Landé g-factor and the tiny band gap of the system. To better evaluate the Berry phase from the SdH oscillations, we consider the low field oscillations ($B < 1$ T) and fit them with the Lifshitz-Kosevich (LK) formula (See Methods). The Berry phase $\Phi_B$ is connected with the fitting phase factor $\gamma$ in the LK equation through $\frac{\Phi_B}{2\pi} = 1/2 - \gamma - \delta$, where $\delta$ is a phase factor associated with the Fermi surface dimensionality. For Dirac fermions in the 2D case, $\delta = 0$ and for the 3D case, $\delta = \pm 1/8$, where the sign depends on the carrier type and the extremal Fermi surface cross-section. In bulk ZrTe$_5$, $\gamma = 0.125$ was measured previously and it was consistent with a $\pi$ Berry's phase, demonstrating Dirac fermions in a 3D Fermi surface [37,38,58]. Similarly, for our unstrained samples, the fitted phase factor is $\gamma = 0.12 \pm 0.01$, as shown in Fig. 4b, and for strain $\epsilon_1$ the fitted $\gamma = 0.13 \pm 0.02$. Indicating the charge carriers responsible for the SdH oscillations for the unstrained and strain $\epsilon_1$ cases are bulk Dirac carriers with a 3D Fermi surface. However, for strain $\epsilon_2$, the fitted phase factor changes significantly to $\gamma = -0.006 \pm 0.040$, which deviates strongly from the 3D case and supports the scenario of 2D Dirac fermions.

Additionally, we perform longitudinal magneto-transport measurements by aligning the sample's *a* axis (parallel to the electrical current direction) to the *B*-field direction. Fig. 4c shows $\rho_{xx}$ vs. *B* at different strains. Without applied strain, $\epsilon_0$, we observe strong SdH oscillations with a frequency of 5.35 T, which corresponds to the anisotropic 3D bulk Fermi surface as observed in previous studies [29,30]. With increasing strain to $\epsilon_1$, we observe that the SdH oscillations vanish, as if the Fermi surface is not 3D anymore. By increasing strain to $\epsilon_2$, $\rho_{xx}$ vs. *B* becomes positive and linear, even up to $B = 9$ T (Fig. 4c). For clarity, we show the longitudinal magnetoresistance, $LMR(B) = (\rho_{xx}(B) - \rho_{xx}(B = 0))/\rho_{xx}(B = 0)$, in Fig. 4d.

For the zero strain case, $\epsilon_0$, the LMR can be well fitted by a quadratic equation $LMR(B) = \eta B^2$, where $\eta$ is a fitting parameter. Under strain $\epsilon_1$, the LMR is reduced by an appreciable amount, showing a competition between a negative and a positive LMR. The LMR clearly shows that it first decreases with an increasing $B$, reaching a minimum, after which, it starts to increase with increasing $B$. This suggests that under strain $\epsilon_1$, an additional scattering mechanism may be relevant (see Methods). The negative LMR in the low field range can be roughly described by a positive magneto-conductance with a quadratic field dependence $\sigma_1(B) \propto B^2$, while the positive LMR above $B = 1.25$ T exhibits a quadratic field dependence $MR(B) \propto B^2$. Without considering the exact type of mechanism, according to Matthiessen's rule (See Methods)[59], the net LMR can be written as $LMR(B) = \frac{1}{\sigma(0)+\eta_1 B^2} - \frac{1}{\sigma(0)} + \eta_2 B^2$, where $\eta_1$, $\eta_2$, and $\sigma(0)$ are fitting parameters. Our data can be well fitted with this equation for $B$ up to 3 T, as shown in Fig. 4d and the inset. The negative LMR up to -7% probed near $B = 0$ may be attributed to the intrinsic chiral anomaly effect as reported in previous studies on the pentatellurides[26,60], especially in the DSM phase which is in good agreement with our assessment that the band gap is closed under strain $\epsilon_1$. By increasing strain to $\epsilon_2$, the LMR can be well fitted by a linear equation $LMR(B) = k|B|$ (as shown in Fig. 4d for B < 2T or Fig. 4c for $B$ up to 9T), where $k$ is a fitting parameter. This linear LMR may be related to the spin polarization of the helical TSS previously observed in other STI materials[61] in agreement with our assessment of the reopening of the band gap under strain $\epsilon_2$ and the dominance of the TSS.

**Discussion and conclusions:**

The strain-induced TPT observed in our experiments can be interpreted as follows. By applying tensile strain along the $c$ axis of HfTe$_5$, initially, the bulk gap is reduced until it is fully closed, and eventually, the band gap reopens with increasing strain. Once the gap reopens the TSS are formed. Causing the transition from the WTI phase to a Dirac semimetal phase and finally an STI phase with dominant TSS. Our first-principles calculations are in good agreement and show that the bulk gap is initially reduced and then increased when increasing tensile strain along the $c$ axis in HfTe$_5$.

As far as we know, signatures of in-situ strain-driven TPTs have been observed only in TaSe$_3$ and ZrTe$_5$ samples by ARPES[33,57,62] and electronic transport[40,41]. In TaSe$_3$, a strain-induced TPT from an STI phase to a trivial semimetal phase was observed by ARPES[57,62]. However, due to the dominant bulk band at the Fermi level[57,62], the appearance or disappearance of the TSS would be difficult to detect from transport measurements. Thus, quantum transport evidence of the TPT in TaSe$_3$ has not been reported yet. In ZrTe$_5$, an ARPES study attempted to observe the TPT by applying strain along the *a* axis[33]. With the maximum compressive strain ($\epsilon_a \sim -0.3\%$). The band gap was reduced, but the STI phase was not reached in the study. There have been two other electronic transport studies focusing on magneto-transport signatures of the strain-tuned TPTs in ZrTe$_5$[40,41]. However, the samples in those studies were dominated by bulk carriers and no transport signatures of the TSS associated with the STI phase were observed. In previous strain studies, the change of the gap was small and an observation of insulating states (reopening of the gap) to confirm the TPT to an STI phase was not observed, which was probably due to the limited compressive strain ($\epsilon_a$ < 1 %) without causing buckling on the sample. On the other hand, in this study, we show transport evidence of a TPT from a WTI phase to an STI phase in HfTe$_5$. Instead of applying compressive strain along the crystallographic *a* axis, we apply tensile strain along the *c* axis with a homemade bending apparatus, Fig. 3a, which allows the application of a much larger strain range. Compared with that of the unstrained $\epsilon_0$ case, the resistivity at the base temperature increases by more than two orders of magnitude with the highest strain $\epsilon_3$ applied. By driving the system deeper into the STI phase, its temperature dependence of resistivity exhibits a semiconducting-like behavior at all temperatures and tends to saturate at low temperatures, which is a key signature of the surface-state dominated transport[42–45]. The dramatic changes in resistivity with the application of strain reveal the changes in the bulk gap across the TPT driven by strain. The remaining finite conductance with the gapped bulk state serves as strong transport evidence of the exposed TSS. Our magneto-transport results at different strains can be well understood in terms of the bulk-dominated transport at small strains and surface-state-dominated transport at high strains. Under a perpendicular magnetic field, the SdH oscillations persist even at high strains when the low-temperature resistivity behaves as an insulator. Detailed analysis of the SdH oscillations phase shift, $\gamma = 1/2 - \frac{\Phi_B}{2\pi} - \delta$, shows the dimensionality nature of the SdH oscillations changes with strain. At zero strain $\epsilon_0$, the SdH oscillations analysis result in $\gamma \sim 1/8$, which supports a 3D Fermi surface in addition to the non-trivial $\pi$ Berry phase of the Dirac electrons ($\delta = -1/8$). At high strain, $\epsilon_2$, the resulting phase shift $\gamma \sim 0$ implies a dimensional phase factor of $\delta = 0$, which is in line with a 2D Fermi surface once the sample is in the STI phase. An alternative explanation is that the system goes through a Lifshitz transition under strain, i.e., the bulk Fermi surface changes from a closed 3D ellipsoid-like to an open 2D cylindrical shape with open orbits along the out-of-plane direction. However, this might not be expected as the interlayer interaction should be stronger, and the momentum transfer should be favored by the compressive strain along the interlayer direction (*b* axis) under the in-plane tensile strain.

The LMR at zero strain, $\epsilon_0$, further supports the 3D Fermi surface of HfTe$_5$. The weak field $B^2$ dependence background LMR and the SdH oscillations can be interpreted well by effective mass anisotropy in the classic limit of the bulk 3D FS and the quantization of electron orbits as it approaches the quantum limit [63]. With a small strain applied, $\epsilon_1$, the negative LMR in the small $B$ range supports the hypothesis of initially closing the bulk gap, and the observation of the chiral anomaly with imbalanced chirality in the intermediate gapless or slightly gapped DSM state[60,64], as it was previously observed in $ZrTe_5$[26,40,58]. It implies that the Dirac gap is closing with the small strain $\epsilon_1$. At higher strain $\epsilon_2$, the SdH oscillations and the negative LMR vanish. Instead, we observe a linear LMR without SdH oscillations. This finding agrees well with the formation of a 2D Fermi surface of Dirac electrons on the top and bottom surfaces with the chemical potential located within or in the vicinity of the bulk gap. The appearance of the compelling linear LMR may be related to the coupling between helically polarized spins of the TSS and the magnetic field, as previously reported in 3D STIs[61], which deserves further investigation.

In summary, we have performed a thorough study on the application of controllable large strain in high-quality $HfTe_5$ samples. We perform DFT calculations and measure the electronic transport properties under strain. The electronic band structure calculations accurately confirm that the $HfTe_5$ ground state is located at the boundary between a WTI and an STI phase. By using a small strain, we prove the $HfTe_5$ samples are initially in the WTI phase. By applying large strain, we observe a strain-driven TPT from the initial WTI phase to a closing (DSM phase) and reopening of the band gap, and finally an STI phase. With a strain of $\sim 4.5\%$, the sample resistivity at $T = 1.5$ K is increased by more than two orders of magnitude. We measured signatures of the topological surface-state-dominated electronic transport in the STI phase of $HfTe_5$ from magneto-transport with both perpendicular and parallel magnetic fields. Our results demonstrate that $HfTe_5$ is an ideal prototype material for studying strain-driven quantum phenomena and has the potential for use in strain-controllable topological spintronic devices.

## Methods

### Crystal growth and structural characterization

The HfTe$_5$ single crystals in this study are grown by the chemical vapor transport (CVT) method. Hf pieces and Te lumps are mixed in stoichiometric ratios before being loaded into a quartz tube (inner diameter 14 mm). A small amount of I$_2$ (100 mg) is added as the transport agent. After being sealed under vacuum, the quartz tube with a length of ~16 cm is placed in a horizontal furnace and the temperature gradient is set to be 510 °C and 460 °C. Belt-like single crystals up to 1 cm in length are accessible after one month of growth. A piece of as-grown crystal was isolated and cut into 1/3 of the length for the single crystal X-ray diffraction measurement. Extreme care was taken while cutting the crystal to minimize straining the crystal edge. The crystal structure is confirmed to be orthorhombic with a space group of *Cmcm*.

### Transport measurements

The electrical measurement is done with the four-probe method. To ensure good contact between electrodes and the sample, 10 nm Cr and 90 nm Au are evaporated on freshly cleaved HfTe$_5$ single crystals under homemade shadow masks in a high vacuum e-beam evaporator (Angstrom Engineering Inc.). Two-component H20E silver epoxy (Epoxy Technology, Inc.) is used to attach 50μm thick Pd-coated copper wires onto the sample and baked at 90 °C for 1 hour in the inert gas atmosphere. The prepared samples for electrical transport measurements are then either measured as "free-standing" samples or pasted on the Ti beam for measurements under bending strain. The samples are cooled down with a helium gas environment in a cryostat equipped with a cryogen-free superconducting magnet (Cryomagnetics, Inc). The temperature is monitored with a Cernox temperature sensor integrated into the sample platform. The electrical measurements are performed by using SR830 lock-in amplifiers (Stanford Research Systems, Inc.). An AC current, typically, of 100 $\mu$A with a frequency of 17.777 Hz is applied along the samples' *a* axis.

### Angle-resolved photoemission spectroscopy (ARPES) measurements

Laboratory-based ARPES experiments were performed on HfTe$_5$ single crystals cleaved in situ at $T$ = 77 K in a base pressure P < 1 × 10$^{-10}$ Torr. Photoelectrons were collected over a 30° solid angle following He I$_\alpha$ ($hv$ = 21.2 eV) excitation using a Scienta R4000 electron energy analyzer. The total energy resolution was $\Delta$E < 12 meV, while an instrumental angular resolution of $\pm$ 0.5° gives rise to a total momentum resolution $\Delta$k < 0.02 Å$^{-1}$ for the photon energy used in these experiments. Calibration of the spectrometer work function was performed on polycrystalline gold at $T$ = 77 K ($\phi$ = 4.345 eV), providing an absolute reference for the Fermi level ($E_F$) as is necessary given the insulating nature of HfTe$_5$ measured from transport in this temperature range.

**DFT Calculations**

Our first-principles calculations are performed with the projector-augmented wave pseudopotentials[65] using Vienna Ab initio Simulation Package[66] code. We have compared four different exchange-correlation functionals: standard Perdew-Burke-Ernzerhof (PBE)[51], PBE-D3 with a van der Waals (vdW) correction[52], optB86b-vdW (a modified vdW-DF functional)[53], and the strongly constrained and appropriately normed (SCAN)[54] with the revised Vydrov-van Voorhis (rVV10)[55]. An energy cutoff of 450 eV and an 8 × 8 × 4 Monkhorst−Pack k-point grid is used[67]. The structure is optimized until the atomic forces are smaller than 0.01 eV/Å. The maximally-localized Wannier functions of HfTe$_5$ are fitted based on the Te-p orbitals by the Wannier90 code[68] and then the topological properties are calculated by the WannierTools[69].

When applying strain along the *c* axis (or *a* axis), the lattice constants *a* (or *c*) and *b* are reduced correspondingly. The Poisson's ratios are calculated from the energy minimization from DFT calculations, shown as follows with Eq. (1) for uniaxial strain along the c-axis and Eq. (2) for uniaxial strain along the a-axis:

$$\begin{cases} \frac{\Delta a}{a} = -0.29\frac{\Delta c}{c} \\ \frac{\Delta b}{b} = -0.12\frac{\Delta c}{c} \end{cases} (1); \quad \begin{cases} \frac{\Delta c}{c} = -0.29\frac{\Delta a}{a} \\ \frac{\Delta b}{b} = -0.05\frac{\Delta a}{a} \end{cases} (2).$$

**Application of bending strain**

An air-annealed titanium beam is placed under two brackets with its center supported by a sapphire bead (Figs. 3b i and ii). As the brackets are driven downward by tightening the screws, the beam will be bent, creating tensile strain along the length direction (defined as *x* axis) on the top surface of the beam. Such a strain can be well approximated by $\epsilon = t/(2R)$, where $\epsilon$ is the strain, *t* is the thickness of the beam and R is the bending radius (See Section IX of the SI). The sample is pasted along the middle line of the beam top surface by insulating EP29LPSP two-component epoxy (Master Bond Inc). A baking process of the epoxy is done inside an Ar-filled glove box by following the instructions given by Master Bond. The sample's *a* axis is aligned perpendicular to the *x* axis. Thus, as the beam is bent, a tensile strain along the *c* axis and a compressive strain along the *a* axis will be applied to the sample. The thin insulating oxide layer on the top surface of the beam produced by the air-annealing process, together with the insulating epoxy, enables us to do electrical transport measurements of samples under strain. A finite element analysis (FEA) simulation of the strain distribution on the beam along the *x* and *y* axes was performed using the Stress Analysis environment in Autodesk Inventor 2023 as depicted in Figs. 3b iii and iv. The beam's dimensions are set to 2 mm ×10 mm × 1 mm (width, length, and thickness), and the Young's modulus and Poisson's ratio are set to 103 GPa and 0.36, respectively. While the Young's modulus of the sapphire bead was set to 345 GPa. Regarding the mesh construction of the assembly, the average element size was 10% of the model size (part-based measure was used) and the minimum element size was 20% of the average value, resulting in a total of 10378 elements and 17302 nodes in the simulation. A force of 290 N was applied downward to each bracket to bend the beam resulting in a 2.3% longitudinal strain at its center, which agrees with the value calculated from measuring the bending radius R.

**Quantum oscillations fitting with the Lifshitz-Kosevich (LK) formula**

Lifshitz-Kosevich (LK) formula:

$$\Delta\rho \propto B^{\lambda} R_T R_D R_S \cos\left[2\pi\left(\frac{F}{B} + \gamma\right)\right] \quad (3)$$

Where $R_T = \alpha T\mu/[B \sinh(\alpha T\mu/B)]$ is the thermal damping term, $R_D = \exp(-\alpha T_D \mu/B)$ is the Dingle damping factor with $T_D$ being the Dingle temperature, and $R_S = \cos(\pi g\mu/2)$ is the spin damping factor. $\mu = \frac{m^*}{m_e}$ is defined as the ratio between cyclotron effective mass m* and the free electron mass $m_e$. $\alpha = 2\pi^2 k_B m_e/(\hbar e)$ is a constant, where $k_B$ is the Boltzmann constant and $\hbar$ is the Planck constant. $g$ is the Landé g-factor. $\lambda$ is a factor depending on the dimensionality (*1/2* and 0 for 3D and 2D cases respectively).

**Matthiessen's rule**

For a single type of carrier, according to Matthiessen's rule,

$$\frac{1}{\tau} = \frac{1}{\tau_1} + \frac{1}{\tau_2} + \cdots \quad (4)$$

where $\tau$ is the total mean free time and $\tau_i$ is the mean free time caused by the $i^{th}$ scattering mechanism. As a result, the net resistivity can be written as the sum of the resistivity introduced by the $i^{th}$ scattering mechanism, *i.e.* $\rho(B) = \rho_1(B) + \rho_2(B) + \ldots$. Different from the additive of conductivity for systems with different conduction channels or different types of charge carriers, here the resistivity caused by different scattering mechanisms experienced by the same carriers is additive. In our case, one scattering mechanism causes positive magneto-conductance, $\rho_1(B) = 1/(\sigma_1(0) + \beta_1 B^2)$, and another scattering mechanism causes positive magneto-resistance, $\rho_2(B) = \rho_2(0) + \beta_2 B^2$, where $\sigma_1(0)$, $\rho_2(0)$, $\beta_1$, and $\beta_2$ are fitting parameters. The total resistivity ($\rho(B) = \rho_1(B) + \rho_2(B)$) can be expressed by $\rho(B) = \frac{1}{\sigma_1(0) + \beta_1 B^2} + \rho_2(0) + \beta_2 B^2$ and the longitudinal magnetoresistance (LMR) equation can be expressed as $LMR(B) = \frac{1}{\sigma(0) + \eta_1 B^2} - \frac{1}{\sigma(0)} + \eta_2 B^2$ (where $\eta_1$, $\eta_2$, and $\sigma(0)$ are fitting parameters) which satisfy the condition LMR (0) = 0. This LMR equation describes our data under strain $\epsilon_1$ very well.

## Acknowledgments

This research was primarily supported by the National Science Foundation Materials Research Science and Engineering Center program through the UC Irvine Center for Complex and Active Materials (DMR-2011967). L.A.J. acknowledges the support from NSF-CAREER (DMR 2146567). M.T.P. and N.S. acknowledge support from the Laboratory Directed Research and Development program of Los Alamos National Laboratory under project number 20230014DR. This work was performed, in part, at the Center for Integrated Nanotechnologies, an Office of Science User Facility operated by the U.S. Department of Energy (DOE) Office of Science. Los Alamos National Laboratory, an affirmative action equal opportunity employer, is managed by Triad National Security, LLC for the U.S. Department of Energy's NNSA, under contract 89233218CNA000001. We acknowledge discussions of the data with Cyprian Lewandowski at Florida State University and Shunqing Shen at the University of Hong Kong. J.L and L.A.J are grateful to Jing Xia and his group at UCI for the initial exploration of using a three-piezo stack strain cell to apply strain on a $HfTe_5$ sample.

## Competing interests

The authors declare no competing interests.

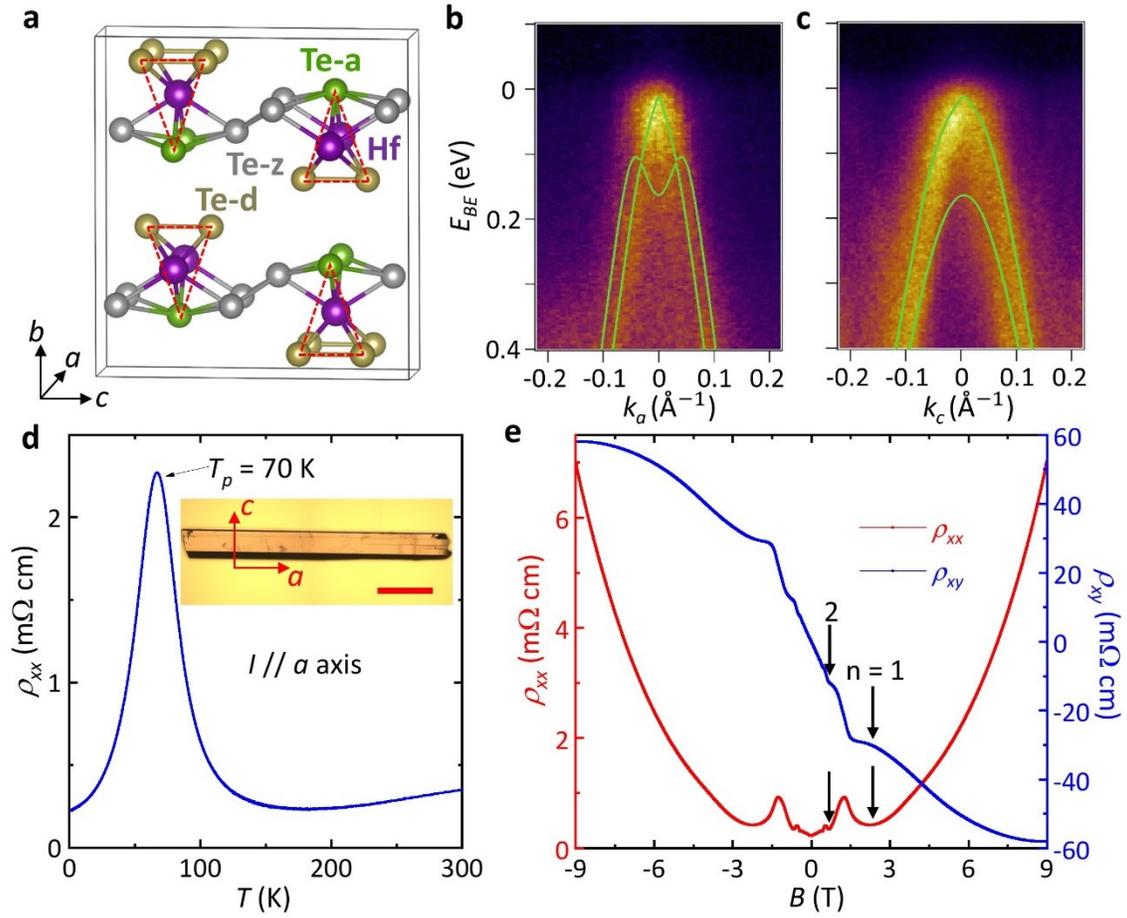

**Figure 1. Characterization of high-quality HfTe$_5$ single crystals. a,** Crystal structure of HfTe$_5$. Te-d, Te-z, and Te-a represent Te atoms at dimer, apical, and zig-zag positions, respectively. **b & c,** ARPES results for the dispersions along X-Γ-X and Y-Γ-Y directions, respectively. The bright green lines are the band dispersions calculated by DFT, which agree well with the ARPES data. **d,** Temperature ($T$) dependence of resistivity ($\rho_{xx}$) for a "free-standing" HfTe$_5$ sample without any external strain applied. Inset: An optical image of a belt-like HfTe$_5$ single crystal, and the scale bar represents 0.2 mm. **e,** Longitudinal ($\rho_{xx}$) and Hall ($\rho_{xy}$) resistivity plotted as a function of magnetic field ($B$) up to 9 T at $T$ = 1.5 K. The arrows mark the Landau level indexes n = 1 and 2.

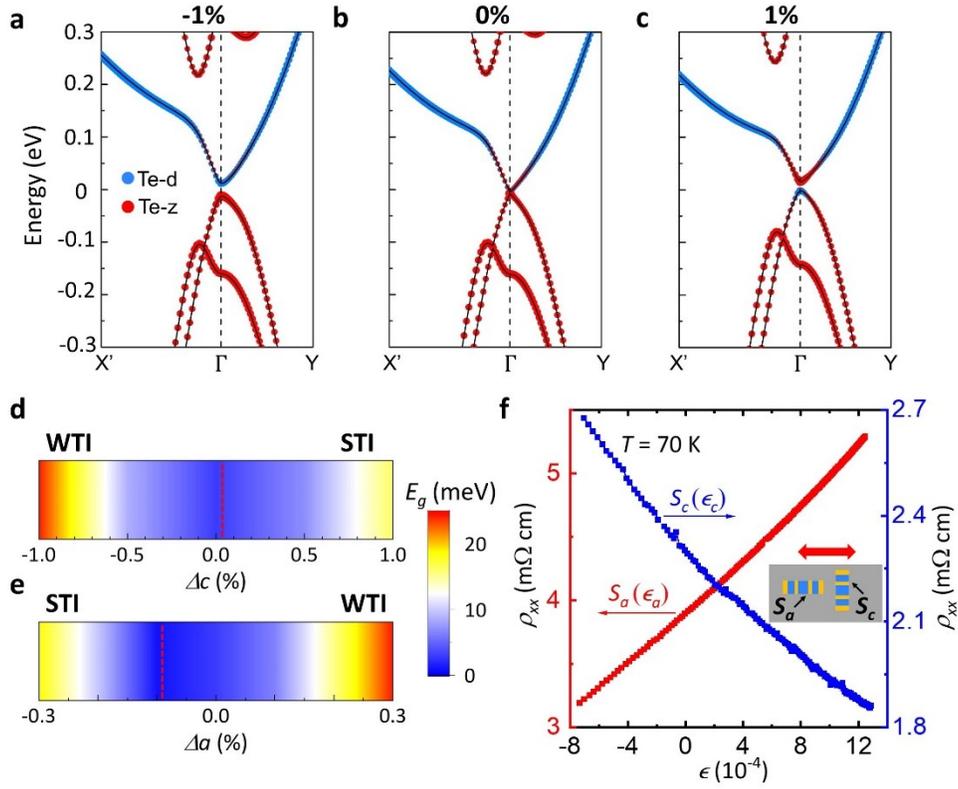

**Figure 2. Topological properties and Topological phase transition under strain in HfTe$_5$. a-c,** Band structure of HfTe$_5$ with different uniaxial strains applied along the axis (-1%, 0%, and 1% for a, b, and c, respectively). A band inversion occurs between the Te-d (blue dots) and Te-z (red dots) orbitals with tensile strain along the c axis. **d & e,** The evolution of the band gap at Γ, with strain applied along c axis in d and along a axis in e. The red dashed lines represent the estimation of the gap closing point. **f,** $\rho_{xx}$ as a function of strain for both Sample $S_a$ and $S_c$ in a double Y-axis plot. Inset: Schematic of sample configurations mounted on a single piezo-stack actuator for applying small strain along either c (Sample $S_c$) or a (Sample $S_a$) axes. The red double-headed arrow represents the poling direction of the piezo-stack actuator.

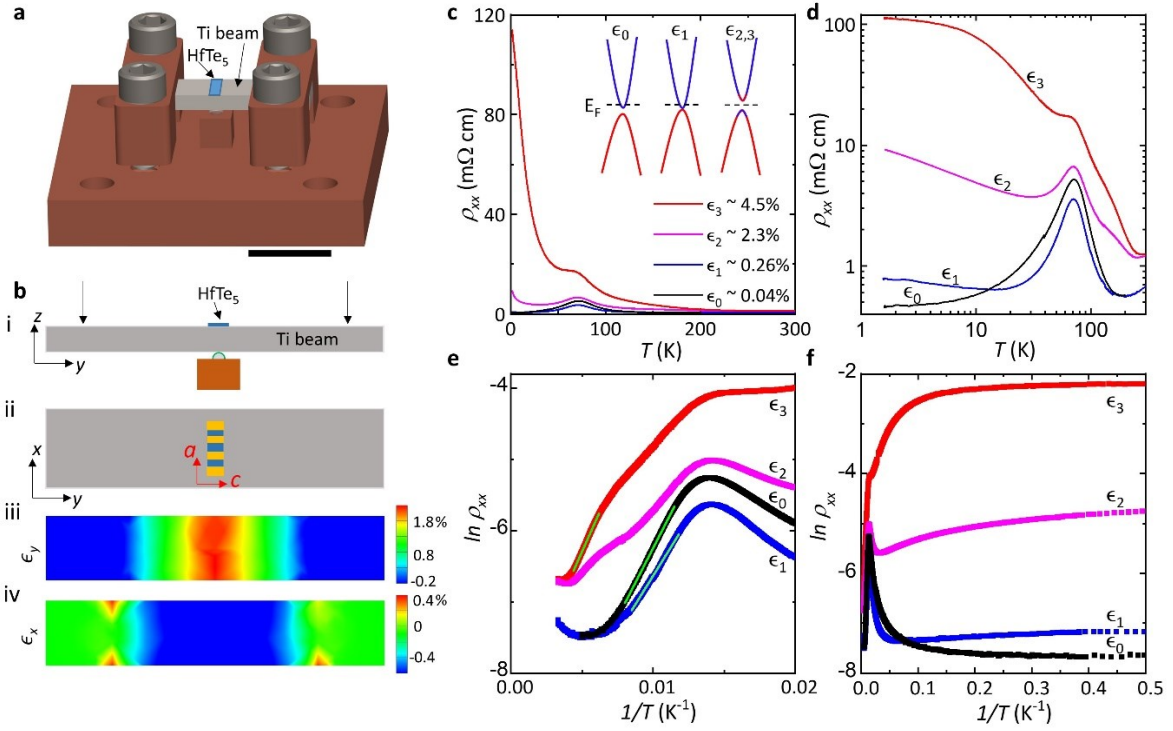

**Figure 3. Tuning electronic transport in HfTe$_5$ by large bending strain. a,** A model of the "bending station" with a sample mounted at the center of the beam's top surface. The scale bar represents 1 cm. **b,** i. Schematic of how the beam is bent. The two black arrows represent the force load. ii. Top view of the sample configuration relative to the beam for applying strain along the sample's *c* axis. iii. and iv. show the strain distribution of $\epsilon_y$ and $\epsilon_x$ on the beam under a moderately high load, which results in a strain of $\epsilon_y$ = 2.2 % near the sample area. **c & d,** $\rho_{xx}$ as a function of temperature (*T*) for the HfTe$_5$ sample under different strains, plotted in log-log scale in d for clarity. Inset of c: Schematics of the Dirac bands around Γ under different strain cases. **e & f,** ln($\rho_{xx}$) *vs.* *1/T* at various strains, plotted for the high-temperature range in e and the low-temperature range in f. The bright straight solid lines in e are the linear Arrhenius fit to extract the thermal activation energy.

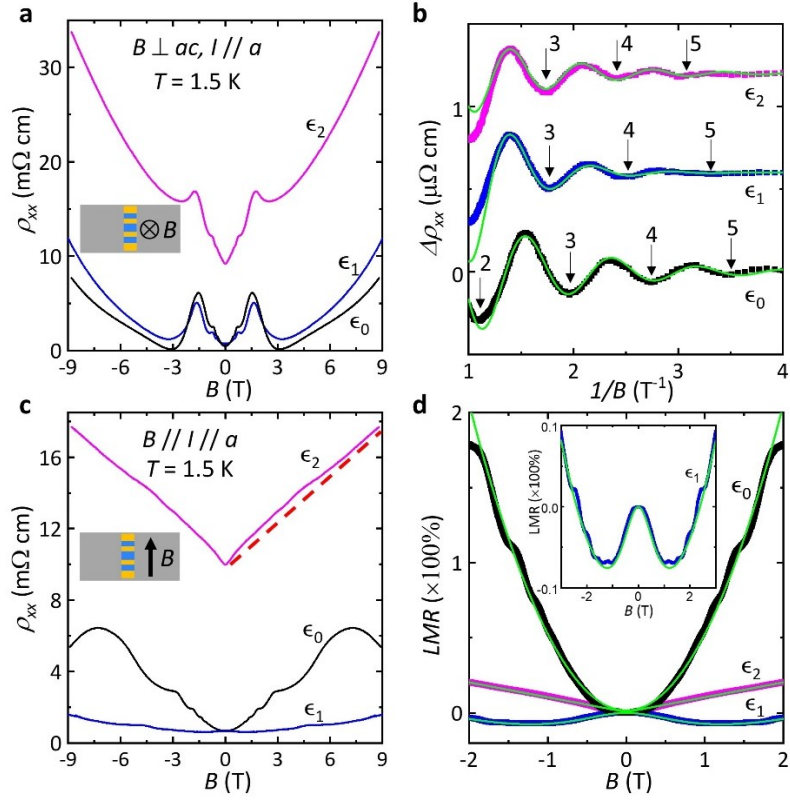

**Figure 4. Magneto-transport in HfTe$_5$ under different strains. a,** $\rho_{xx}$ as a function of perpendicular magnetic field ($B$) measured under different strains. Inset: Schematic of the measurement setup. **b,** Oscillatory part of the resistivity ($\Delta\rho_{xx}$) after background subtraction, plotted vs. $1/B$ in the field range from 0.25 T to 1 T. $\Delta\rho_{xx}$ is vertically shifted for clarity for strains $\epsilon_1$ and $\epsilon_2$. The bright solid green lines represent the LK fit (See Methods). The arrows mark the positions of integer Landau levels. **c,** $\rho_{xx}$ as a function of parallel magnetic field ($B$) at different strains, measured with $B \,//\, I$ (for this we rotate the sample and aligned the crystal *a* axis to $B$). The red dashed line is a guide for the linear magnetic field dependence of $\rho_{xx}$ vs. $B$ measured under strain $\epsilon_2$. Inset: Schematic of the measurement setup. **d,** Longitudinal magnetoresistance (LMR) vs. $B$ (up to 2 T). The bright solid green lines represent the fittings with different equations, $LMR(B) = \eta B^2$ for $\epsilon_0$, $LMR(B) = \frac{1}{\sigma(0)+\eta_1 B^2} - \frac{1}{\sigma(0)} + \eta_2 B^2$ for $\epsilon_1$, and $LMR(B) = k|B|$ for $\epsilon_2$. Where $\eta_1$, $\eta_2$, $\sigma(0)$, and $k$ are fitting parameters. Inset: Zoomed-in plot of LMR vs. $B$ (up to 3 T) for strain $\epsilon_1$. All the measurements were performed at $T = 1.5$ K.